\documentclass[1p]{elsarticle}

\pdfoutput=1 

\usepackage{lineno,hyperref}
\graphicspath{ {Figuras/} }
\usepackage[export]{adjustbox}
\usepackage{float}
\usepackage{amsmath}
\usepackage{xcolor}
\usepackage{multirow}

\usepackage{vmargin}
\setpapersize{A4}
\setmargins{2.5cm}       
{1cm}                    
{16cm}                   
{25cm}                   
{10pt}                   
{1cm}                    
{0pt}                    
{1.5cm}                  

\makeatletter
\def\@author#1{\g@addto@macro\elsauthors{\normalsize%
    \def\baselinestretch{1}%
    \upshape\authorsep#1\unskip\textsuperscript{%
      \ifx\@fnmark\@empty\else\unskip\sep\@fnmark\let\sep=,\fi
      \ifx\@corref\@empty\else\unskip\sep\@corref\let\sep=,\fi
      }%
    \def\authorsep{\unskip,\space}%
    \global\let\@fnmark\@empty
    \global\let\@corref\@empty  
    \global\let\sep\@empty}%
    \@eadauthor={#1}
}
\makeatother

\journal{ArXiv.org}

\begin{document}


\begin{frontmatter}

\title{Estimation of lateral track irregularity through Kalman filtering techniques}


\author{Sergio Mu\~noz\fnref{Sergio_address}\corref{cor1}}
\author{Javier Ros\fnref{Javier_address}}
\author{Jos\'e Luis Escalona\fnref{Jose_address}}

\cortext[cor1]{Corresponding author: sergiomunoz@us.es}

\address[Sergio_address]{Department of Materials and Transportation Engineering, University of Seville, Spain}
\address[Javier_address]{Department of Mechanical Engineering, Public University of Navarre, Spain}
\address[Jose_address]{Department of Mechanical Engineering, University of Seville, Spain}


\begin{abstract}

The aim of this work is to develop a model-based methodology for monitoring lateral track irregularities based on the use of inertial sensors mounted on an in-service train. To this end, a gyroscope is used to measure the wheelset yaw angular velocity and two accelerometers are used to measure lateral acceleration of the wheelset and the bogie frame. Using a highly simplified linear bogie model that is able to capture the most relevant dynamic behaviour allows for the set-up of a very efficient Kalman-based monitoring strategy. The behaviour of the designed filter is assessed through the use of a detailed multibody model of an in-service vehicle running on a straight track with realistic irregularities. The model output is used to generate virtual measurements that are subsequently used to run the filter and validate the proposed estimator. In addition, the equivalent parameters of the simplified model are identified based on these simulations. In order to prove the robustness of the proposed technique, a systematic parametric analysis has been performed. The results obtained with the proposed method are promising, showing high accuracy and robustness for monitoring lateral alignment on straight tracks, with a very low computational cost. 

\end{abstract}


\begin{keyword}
\texttt{rail vehicle dynamics; track irregularity; kalman filtering}
\end{keyword}

\end{frontmatter}


\section{Introduction}

The main function of railway tracks is the correct guidance of the vehicle, without compromising its stability. These two requirements, guidance and stability, are usually in conflict and a balance must be achieved \citep{Wickens2003}. Any deviation from the ideal track geometry can excite unwanted vehicle dynamic responses, leading to poor ride quality or, possibly, to safety problems. These deviations, called track irregularities, are usually described using four variables \citep{Esveld2001}: 1) \textit{track gauge variation} and 2) \textit{lateral alignment} for horizontal deviations, and 3) \textit{cross-level} and 4) \textit{vertical profile} for vertical ones. In the European Union, for instance, the Standard EN13848 \cite{Standars_EN13848} is used to define the acceptable limit levels for track irregularities according to their wavelength in three different ranges: $ D_1=[3, 25]$ m , $D_2 =[25, 70]$ m and  $ D_3=[70, 200] $ m. Consequently, when evaluating the quality of the track geometry, irregularities should be analysed taking into account both these wavelength ranges, as well as the maximum allowed forward velocity of the vehicle.

It is essential that the maintenance of the railway tracks meets the appropriate standards of quality for both ride safety and passenger comfort in the vehicle. In this respect, continuous monitoring of track geometry is usually carried out through the use of track recording vehicles (TRV), which provide an accurate measurement of irregularities using different sets of optical, laser or inertial sensors. However, the use of these dedicated trains with sophisticated measuring devices is complex and very expensive. As an alternative, the development of inexpensive measuring systems to be used on in-service vehicles for continuous monitoring of track conditions seems very attractive. Consequently, simple and robust measuring systems, combined with the development of dynamic model-based filtering techniques, are required to achieve an accurate estimation of the track geometry. Some work in this direction has already been carried out. An extensive review on the perspectives on the use of in-service vehicles for the monitoring of railway tracks can be found in \cite{Weston2015}. Even though their conclusions are promising, most of the works referred to have no experimental validation, being academic in nature. The use of model-based Bayesian filtering techniques, such as Kalman filtering, are pointed out as the most promising approaches for monitoring track geometry.

Regarding the monitoring of vertical irregularities, the simplest methods consist of the integration of accelerometer and gyroscope derived signals to obtain the absolute position of the wheelset and hence the track geometry. For example, in \cite{ Weston2007}, the vertical profile of the track is estimated by integration of the vertical curvature of the track centre-line, which is derived from the pitch-rate gyroscope sensors mounted on the bogie frame of an in-service vehicle. However, the integration of measured signals leads to low accuracy and a drift in the obtained results. High-pass filtering alleviates the drift problem at the expense of losing information in the low frequency range. In \cite{Escalona2016}, vertical track irregularities are estimated through a sensor fusion algorithm based on complementary filters: the signals from an accelerometer and a gyroscope installed on a bogie are used and the estimation of low frequency irregularity relies on the data from the gyroscope while the high frequency irregularity relies on the data from the accelerometer. Fairly good results are obtained in the estimations, although a loss of accuracy is shown in the case of variable forward velocity of the vehicle. In \cite{Ros2015} and \cite{Tsunashima2014}, vertical track irregularities are identified through Kalman filter-based techniques, using a kinematic and a dynamic model, respectively. Both works result in relatively acceptable accuracy in the estimated irregularities using an accelerometer and a gyroscope.

Regarding lateral irregularities, their estimation is much more difficult and challenging, since the lateral displacement of the wheelset depends not only on the lateral irregularities of the track but also on the lateral sliding of the wheelset relative to the track, which is related to the creep force dynamics. Furthermore, lateral irregularities, especially \textit{lateral alignment}, are shown to be much more influential in the dynamic behaviour of the vehicle than vertical ones. In \cite{ Weston2007b}, the lateral alignment of the track is estimated through integration of lateral curvature of the track centre-line, using a procedure analogous to the one used in \cite{ Weston2007}. In addition to the problems of drift and low smoothness related to numerical integration, the proposed method cannot take into account the lateral displacement of the wheelset relative to the track. In \cite{Lee2012}, a Kalman filter is used as a naive integrator of the lateral acceleration of the wheelset to obtain the lateral displacements and, subsequently, a set of compensation filters are used in the corresponding wavelength bands to correct these predictions from the lateral displacements of the wheelset. A model-based \textit{unknown input identification filter} is used in \cite{Wei2012}. Here, a linearized lateral dynamic model of a bogie with two wheelsets is used. In this work, the use of $H$-infinity theory in order to maximise the sensitivity of the lateral displacement of the wheelset is of note, as is the robustness of the disturbances and system inputs to the displacement estimation error. More recently, in \cite{DeRosa2019}, the authors propose three different model-based methods to estimate both lateral track alignment and cross-level irregularities: 1) pseudo-inversion of the vehicle’s frequency response function (FRF) matrix, 2)  \textit{unknown input estimation} using a deterministic observer and 3) \textit{unknown input estimation} using a linear Kalman filter as a stochastic observer. They use a very complex linear dynamic model of a railway vehicle composed of one car body, two bogies and four wheelsets. In the proposed model and with 17 degrees of freedom, the relative motion between the wheelset and the track has been taken into account, considering the effect of the creepage forces acting at wheel-rail contact. The proposed methodologies have been validated through the use of numerical experiments based on a rich non-linear multibody model. Quite good results in the estimation are obtained with all three methods, especially with the Kalman filter approach. The main drawback of these methods is the complexity of the dynamic model used (17 degrees of freedom) and the high number of the sensors to be installed on the vehicle (36 accelerometers). It is noticeable that this is one of the few references in which real data are used for validation: the FRF approach has been run using real measurement data from a track recording vehicle (TRV). Even though the results are promising, a degradation in performance is shown in comparison with the validation using virtual sensor synthesised data.

Even though there are several references dealing with the estimation of track irregularities, the published literature on the estimation of lateral alignment is relatively scarce and only focuses on tangent (straight) line track segments. Furthermore, a considerable number of works are rather obscure, with important details omitted, making it impossible to reproduce the results. On the basis of the works published by different authors, the best results are achieved by model-based Bayesian filtering methods, such as Kalman filtering, combining dynamic models which include creep contact forces with experimental information from sensors (gyroscopes and accelerometers). All the published works are based on linear dynamics models. In this regard, the use of more complex models \citep{DeRosa2019} does not seem to outperform the simplest ones \cite{Wei2012}. Despite several authors having demonstrated promising results, there appears to be a lack of profound analysis of these results. On the one hand, the results obtained should be thoroughly analysed in the different wavelength ranges, according to the standards. On the other hand, there is a need for a systematic analysis to test the accuracy and robustness of the proposed technique when there is some kind of uncertainty in the system parameters or in the vehicle running conditions. Finally, one remarkable inadequacy of most works in literature is the lack of rigour in the validation of the proposed estimation technique. The validation procedure is usually performed through the use of the same simulation model used by the estimator, making it unrealistic and lacking in critical interest. Furthermore, in works in which a simplified linear model is used by the estimator and a more complex model is used for validation purposes, the identification of the parameters of the linear model is not clear. This is an important issue to deal with, as the accurate identification of these parameters is essential for the good performance of the derived filter, especially when very simplified linear models are used by the estimator.

In this work, a model-based Kalman filtering technique is proposed for monitoring lateral alignment from the measurements of inertial sensors mounted on an in-service vehicle running on a straight track segment. The railway vehicle used in this work consists of four wheelsets, two bogie frames and a car body. The main contribution of this study is the use of a highly simplified linear dynamic model of the vehicle to perform a classical linear Kalman filter for monitoring the lateral alignment of the track. This simplified dynamic model is based on the lateral dynamics of a single wheelset with two generalised coordinates (lateral displacement and yaw rotation) and a suspended frame with only one generalised coordinate (lateral displacement). Such a simplified dynamic model needs to be precisely validated through an accurate identification of the equivalent parameters, which is essential for the good performance of the estimator. A full multibody model of the vehicle is used to generate virtual measurements. These synthesised data are used for two different purposes: firstly, to identify the equivalent parameters of the simplified dynamic model and secondly, to evaluate the estimation error and validate the proposed Kalman filter estimator. Finally, in order to test the robustness of the proposed technique, a systematic parametric analysis has been performed, evaluating the influence that the uncertainty of different parameters and running conditions could have on the estimation error.

The paper is organised as follows: Section \ref{Sec:Est_technique} presents the estimation technique used in this work. In Section \ref{Sec:Dynamic_model}, the dynamic modelling is presented, as well as the equivalent parameter identification procedure. Section \ref{Sec:KF} details the Kalman filter algorithm used in this work. In Section \ref{Sec:Results}, the results of the track alignment estimation are presented and discussed, and a robustness analysis is performed. Finally, Section \ref{Sec:Conclusions} provides the conclusions and summary.

\section{Estimation technique} \label{Sec:Est_technique}
  
A model-based numeric procedure has been developed for the estimation of the lateral alignment of the track from the measurements from inertial sensors mounted on an in-service vehicle. The estimation procedure has been performed under the following assumptions: the track to be measured is a straight segment with no gauge variation, the wheelset has a conical profile and there is no flange contact. All these requirements are usually fulfilled. However, the proposed method should be extended to more general conditions in future works.

The proposed estimation technique is based on the Kalman filtering method, using the measurements from an accelerometer and a gyroscope mounted on the axle-box of the wheelset and an accelerometer mounted in the bogie frame of the vehicle. To develop this method and analyse its performance, two different models are needed. First, the \textit{Complete Simulation Model} (CM), a complete and detailed model of the vehicle used to generate the synthetic sensor data to be used as an input in the Kalman filter. This complete and detailed model will have the function of validating the estimation process. Second, the \textit{Simplified Estimator Design Model} (SM) to be used by the Kalman filter for model equations. This is a simplified dynamic model of the vehicle that must be able to properly reproduce the dynamic behaviour of the wheelset but is simple enough to reduce the computational load of the model-based observer. In this model, the wheelset-track relative motion is taken into account assuming creep forces at the wheel-rail contact, following Kalker’s linear theory \citep{Kalker1982}.

The railway vehicle used in this work, the ML95 vehicle operated by the Lisbon subway and described in \citep{Pombo2004}, consists of four wheelsets, two bogie frames and a car body. Since the CM considers arbitrary-geometry tracks including rail centre line irregularities, these will be generated and included in the simulations. With the use of the CM, the simulation of the vehicle is carried out and the synthetic data of the virtual sensors are generated, to be used as input in the Kalman filter estimator.
 
The results obtained with the proposed method are analysed in the different wavelength ranges defined in the standards. Furthermore, the efficiency of the proposed estimator is proven. Due to the simplicity of the SM, a very low computational cost is required, making the proposed method especially appropriate for real-time applications. Finally, to complete the study and prove the robustness of the proposed technique, a systematic parametric analysis has been performed. Therefore, the influence of the uncertainty of different parameters and running conditions (sensor noise, vertical irregularity, conicity uncertainty and Kalker’s coefficients uncertainty) on the estimator results has been analysed.

\section{Dynamic modeling} \label{Sec:Dynamic_model}

With the aim of validating the proposed estimation technique, the employment of the CM for the generation of synthesised data is of crucial importance. Only with a detailed and feasible CM, much more accurate than the SM, will the validation procedure be realistic. Otherwise, using the same or a very close simulation model and estimator design model would make the validation procedure self-referential and it would have no critical interest. In this section, the definition of track irregularities according to the standards is presented first; then the CM of the vehicle used for the generation of synthesised data will be presented; and finally the SM to be used in the Kalman filter will be introduced.

\subsection{Definition of track irregularities}  \label{SubSec:Definition_Irreg}

\begin{figure}[H]
	\includegraphics[width=0.6\textwidth,center]{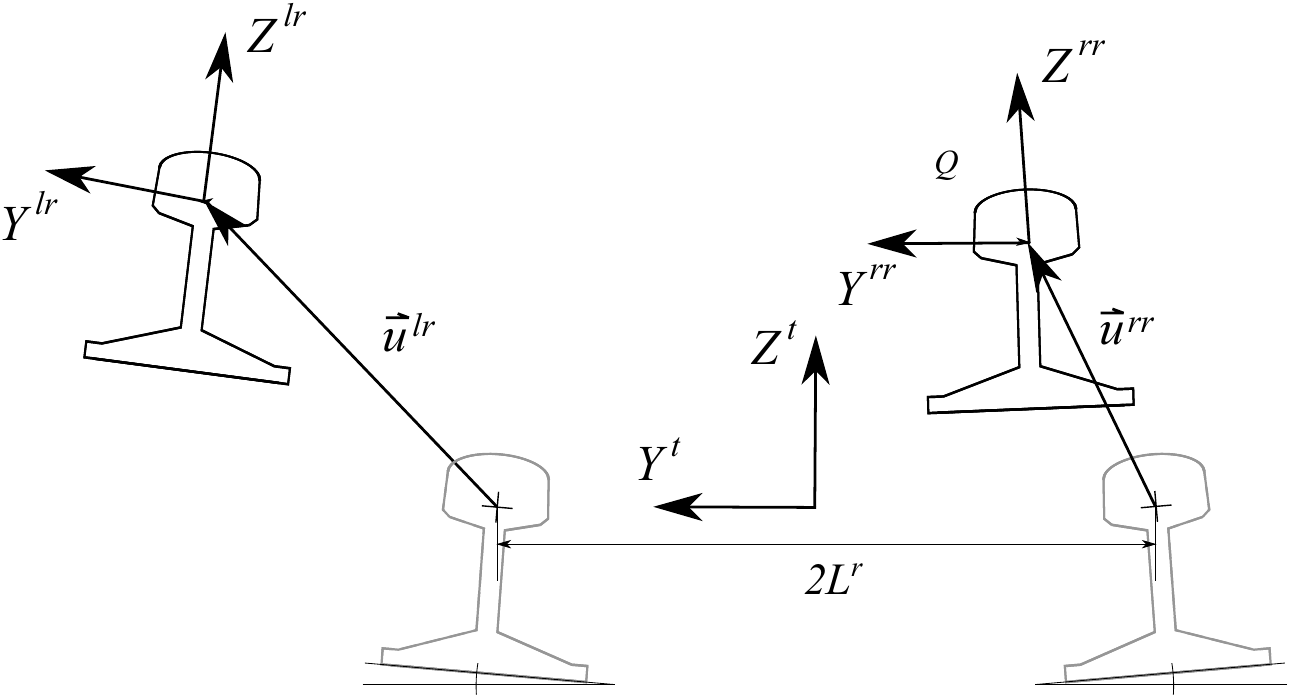}
  	\caption{Definition of track irregularities}
  	\label{fig:Irreg_definition}
\end{figure}

The lateral and vertical irregularities of a track are usually defined in the railway industry by four well-known irregularities variables: track gauge variation, $\xi_{g}$, lateral alignment, $\xi_{a}$, cross-level, $\xi_{cl}$, and vertical profile, $\xi_{vp}$. These variables are defined as follows:

\begin{equation} \label{eq:irregularities}
\begin{split}
\xi_{g} = (u_{y}^{lr}-u_{y}^{rr}),   \qquad \xi_{a} = (u_{y}^{lr}+u_{y}^{rr})/2 \\
\xi_{cl} = (u_{z}^{lr}-u_{z}^{rr}),  \qquad \xi_{vp} = (u_{z}^{lr}+u_{z}^{rr})/2
\end{split}
\end{equation}

where $u_{y}^{lr}$, $u_{y}^{rr}$, $u_{z}^{lr}$ and $u_{z}^{rr}$ are lateral ('$y$') and vertical ('$z$') deviation of the left ('$lr$') and right ('$rr$') rail cross-section from their ideal position, see Fig. \ref{fig:Irreg_definition}.

\subsection{Complete simulation model (CM)}  \label{SubSec:Complete_model}

The CM was presented by the authors in \cite{Munoz2019}. A brief description of the model is given next. The CM is a general model for railway vehicles running on tracks with arbitrary geometry, including irregularities. Because the CM is developed for industrial applications, the number of required parameters for the model is minimised. The CM is general, complete and computationally efficient due to the following features:

\begin{enumerate}
  \item It is based on the use of track-relative unconstrained coordinates. Generalized coordinates are separated into vertical coordinates and lateral coordinates. Bodies are separated into   \emph{wheelsets} and \emph{non-wheelset} bodies.
  \item Kinematic linearization (small-angles assumption) and dynamic linearization of inertia and suspension generalized forces is performed.
  \item It considers weakly coupled vertical and lateral dynamics of the vehicle.
  \item Wheel-rail contact interaction is based on the equivalent conicity concept, the \emph{knife-edge contact} assumption and Kalker’s linear creep theory. Flange contact and two-point contact scenario can be simulated.
  \item Equations of motion are obtained using symbolic computations. The computation of generalized forces is optimized using symbolic computation techniques.
\end{enumerate}

The equations of motion are given by:

\begin{equation} \label{eq:Eq_motion_CM_Vertical}
\mathbf{M}_{V}^{nw} \; \mathbf{\ddot{q}}_{V}^{nw} \; 
+ \mathbf{C}_{V}^{s,nw}  \mathbf{\dot{q}}_{V}^{nw} 
+ \mathbf{K}_{V}^{s,nw}  \mathbf{q}_{V}^{nw} =
\mathbf{Q}_{V}^{ForIn}
- \mathbf{C}_{V}^{s,w}  \mathbf{\dot{q}}_{V}^{w} 
- \mathbf{K}_{V}^{s,w}  \mathbf{q}_{V}^{w} 
+ \mathbf{Q}_{V}^{grav}
+ \mathbf{Q}_{V0}^{s}
\end{equation}

\begin{equation} \label{eq:Eq_motion_CM_Lateral}
\mathbf{M}_{L} \; \mathbf{\ddot{q}}_{L} \; 
+ [\mathbf{C}_{L}^{s} + \mathbf{C}_{L}^{c}] \mathbf{\dot{q}}_{L} 
+ [\mathbf{K}_{L}^{s} + \mathbf{K}_{L}^{c}] \mathbf{q}_{L} =
\mathbf{Q}_{L}^{ForIn}
+ \mathbf{Q}_{L0}^{s}
+ \mathbf{Q}_{L0}^{c}
+ \mathbf{Q}_{L}^{grav}
\end{equation}

where Eq. \ref{eq:Eq_motion_CM_Vertical} includes the equations of motion for the vertical dynamics; $\mathbf{q}_{V}^{nw}$ being the vertical coordinates of the non-wheelset bodies and $\mathbf{q}_{V}^{w}$ the vertical coordinates of the wheelsets (given by the track vertical geometry). In this equation $\mathbf{M}_{V}^{nw}$, $\mathbf{C}_{V}^{nw}$ and $\mathbf{K}_{V}^{nw}$ are the constant mass, suspension damping and suspension stiffness matrices associated with the vertical dynamics; $\mathbf{C}_{V}^{s,w}$ and $\mathbf{K}_{V}^{s,w}$ are the constant suspension damping and suspension stiffness matrices associated with the wheelset vertical coordinates;  $\mathbf{Q}_{V}^{ForIn}$ is the vector of generalised forces due to the forward motion in the vertical direction; $\mathbf{Q}_{V}^{grav}$ is the generalised gravity force vector; and $\mathbf{Q}_{V0}^{s}$ contains the constant terms that appear in the generalised suspension forces. Equation \ref{eq:Eq_motion_CM_Lateral} includes the equations of motion for the lateral dynamics, $\mathbf{q}_{L}$ being the lateral coordinates of the vehicle bodies. In this equation $\mathbf{M}_{L}$, $\mathbf{C}_{L}^{s}$ and $\mathbf{K}_{L}^{s}$ are the constant mass, suspension damping and suspension stiffness matrices associated with the lateral dynamics, respectively;  $\mathbf{C}_{L}^{c}$ and $\mathbf{K}_{L}^{c}$ are damping and suspension matrices associated with the contact forces acting on the wheelset in the lateral direction; $\mathbf{Q}_{L}^{ForIn}$ is the vector of generalised inertia forces due to the forward motion; the vectors $\mathbf{Q}_{L0}^{s}$ and $\mathbf{Q}_{L0}^{c}$ contain the terms that appear in the generalised suspension forces and in the generalised contact forces, respectively, when the lateral coordinates and velocities are zero; and $\mathbf{Q}_{L}^{grav}$ is the vector of generalised gravity forces in the lateral direction.

\subsection{Simplified estimator design model (SM)} \label{SubSec:Simplified_model}

\begin{figure}[h]
	\includegraphics[width=0.85\textwidth,center]{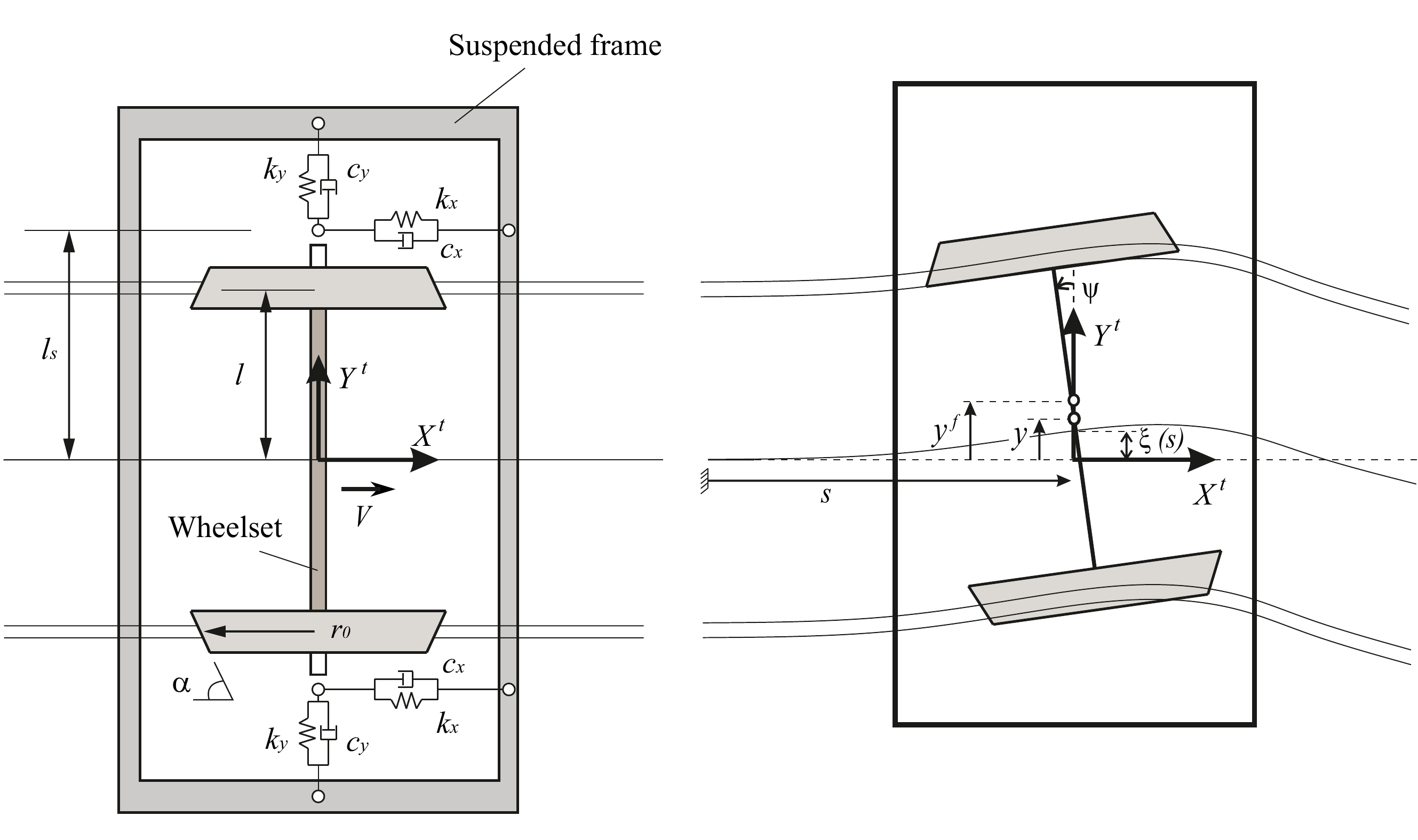}
  	\caption{Plan view of Simplified Model}
  	\label{fig:Simplified_model}
\end{figure}

With the aim of estimating the lateral irregularities of the track through the Kalman filter, a very simplified dynamic model of the vehicle has been used. A schematic plan view of the SM is presented in Fig. \ref{fig:Simplified_model}: the different bodies and elements of the SM are presented in the left drawing, while an arbitrary position of the system is presented in the right one. The SM only models one wheelset and a suspended frame, representing the dynamic interaction of the wheelset with the rest of the vehicle. The SM uses two generalised coordinates for the wheelset (lateral displacement, $y$, and yaw rotation, $\psi$), and one generalised coordinate for the suspended frame (lateral displacement, $y^{f}$). All generalised coordinates are referred to a \textit{Track Frame} $\langle X^{t},Y^{t} \rangle$, that moves along the irregularity-free track centre line with the same forward velocity as the vehicle. Both bodies, the wheelset and the suspended frame, are connected by longitudinal and lateral primary suspension elements.

The set of generalized coordinates of the simplified vehicle model is therefore:

\begin{equation} \label{eq:Gen_coordinates}
\mathbf{q} =\left [ y \; \; \; \psi \; \; \; y^{f}\right ]^{T}
\end{equation}

The equations of motion of the lateral dynamics associated with this model are: 

\begin{equation} \label{eq:Eq_motion}
\mathbf{M} \; \mathbf{\ddot{q}} \; 
+ \; \left \lfloor \mathbf{C}_{s} + \mathbf{C}_{c} \right \rfloor \; \mathbf{\dot{q}} 
+ \; \left \lfloor \mathbf{K}_{s} + \mathbf{K}_{c} \right \rfloor \; \mathbf{q}  =
\mathbf{Q}_{c,0}
\end{equation}

where $\mathbf{M}$, $\mathbf{C}_{s}$ and $\mathbf{K}_{s}$ are the constant mass, suspension damping and suspension stiffness matrices associated with the lateral dynamics, respectively; $\mathbf{C}_{c}$ and $\mathbf{K}_{c}$ are damping and suspension matrices associated with the contact forces acting on the wheelset in the lateral direction; and the vector $\mathbf{Q}_{c,0}$ contains the terms that appear in the generalized contact forces when the lateral coordinates and velocities are zero. These matrices and vectors are obtained using symbolic computation: Lagrange Equations are used to find the inertia, elastic and damping generalised forces, while the principle of virtual work is used to find the contact tangential generalised forces. Wheel-rail contact at the treads is modelled with the \textit{knife-edge contact constraints}, and tread tangential contact forces are calculated following the Kalker’s creep linear theory, as in the CM described in \citep{Munoz2019}. The calculation of the flange contact forces is excluded in this study, according to the assumption of no flange contact.
 
The complete set of parameters that characterises the SM is given by the vector:

\begin{equation} \label{eq:Parameter_vector}
\mathbf{p} =\left [ m \; \; \; I \; \; \; l \; \; \;  l_{s} \; \; \; \alpha \; \; \; r_{0}  \; \; \; m_{f} \; \; \; f_{11}  \; \; \; f_{22} \; \; \; f_{23} \; \; \; f_{33}  \; \; \; k_{x} \; \; \; c_{x}  \; \; \; k_{y} \; \; \; c_{y}\right ]^{T}
\end{equation}

where $m$, $I$ and $l$ are the mass, the yaw moment of inertia and half-width of the wheelset; $l_{s}$ the length to the primary suspension; $\alpha$ the nominal conicity of the wheel tread; $r_{0}$ the rolling radius of the wheels when the wheelset is centered on the track; $m_{f}$ the mass of the suspended frame; $f_{ij}$ the Kalker's linear creep coefficients (that are assumed to be constant); and $k_{x}$, $c_{x}$, $k_{y}$ and $c_{y}$ the parameters of the primary suspension.

The longitudinal position of the vehicle along the track is defined by the arc-length coordinate $s$, that is assumed to be prescribed. Out of the four irregularities defined in Eq. \eqref{eq:irregularities}, only the lateral alignment is considered in the SM, hereinafter referred to as $\xi$. Since the gauge variation is assumed to be zero in this work, the lateral alignment is defined as $\xi$ = $\xi_{a}$ = $u_{y}^{lr}$ = $u_{y}^{rr}$. This lateral alignment is a function of the arc-length coordinate, $\xi=\xi(s)$. The considerable simplification of the vehicle model proposed with the SM should be sufficient to describe the stability and guidance dynamics of the wheelset as a function of the lateral track irregularity. This model will be valid for our purpose only if it can adequately describe the lateral dynamic response of the wheelset running on a straight track with irregularities. Therefore, before using the proposed SM in the Kalman filter, the model has to be identified (in the next section).

\subsection{Identification of equivalent parameters}  \label{SubSec:Param_identfication}

The validation of the SM used by the estimator is crucial for the good performance of the Kalman filter. It is important to note that there is a significant set of simplifications made in the SM. First, the vehicle is modelled just by a single wheelset with two degree of freedom ($y$ and $\psi$), connected to a suspended frame with only one degree of freedom ($y^{f}$). Second, the mass of the suspended frame, which represents the effect of the rest of the train, has constrained yaw rotation. Third, only lateral irregularities can be included in the SM.
The values of all the equivalent parameters of the simplified model, $\mathbf{p}$, given in Eq. \eqref{eq:Parameter_vector}, must be identified to find similar dynamic behaviour of both models (CM and SM). Part of this set of parameters can be directly obtained from the real train: the first six parameters ($m , I , l , l_{s}, \alpha, r_{0}$) correspond to inertial and geometrical properties of the wheelset. The seventh parameter, the mass of the suspended frame ($m_{f}$), can be approximately calculated as the mass supported by the wheelset in a static equilibrium position: that is, a quarter of the mass of the car body plus half the mass of one bogie frame. The next four parameters ($f_{11}, f_{22}, f_{23}, f_{33}$) are the creep coefficients, which depend on normal contact force, the size and shape of the contact patch and the elastic properties of the bodies in contact. In this work, the creep coefficients are assumed to be constant and their values have been calculated following Kalker’s creep linear theory \cite{Kalker1982}, considering the magnitude of the normal contact force as the weight supported by each wheel at the static equilibrium position. Finally, the last four parameters ($k_{x}$, $c_{x}$, $k_{y}$ and $c_{y}$) correspond to the properties of the primary suspension of the SM, which cannot be directly taken from the real suspension elements of the train. These last parameters are more difficult to obtain and, consequently, must be identified by optimisation. In conclusion, the set of parameters can be divided into two subsets: the subset of parameters that can be directly obtained from the real train, $\mathbf{p}_{fix}$, and the subset of parameters to be identified by optimisation, $\mathbf{p}_{opt}$:

\begin{equation} \label{eq:Parameter_vector_divided}
\begin{split}
& \mathbf{p} = \left [\mathbf{p}_{fix}^{T} \; \; \; \mathbf{p}_{opt}^{T} \right ]^{T}\\
& \mathbf{p}_{fix} =\left [ m \; \; \; I \; \; \; l \; \; \;  l_{s} \; \; \; \alpha \; \; \; r_{0}  \; \; \; m_{f} \; \; \; f_{11}  \; \; \; f_{22} \; \; \; f_{23} \; \; \; f_{33} \right ]^{T} \\
& \mathbf{p}_{opt} =\left [ k_{x} \; \; \; c_{x}  \; \; \; k_{y} \; \; \; c_{y}\right ]^{T}
\end{split}
\end{equation}

In order to identify $\mathbf{p}_{opt}$, the simulation of the dynamics of the complete ML95 vehicle \citep{Pombo2004} has been carried out using the CM. Previously, track irregularities had been generated to be included in simulations. For this simulation, the real parameters of the vehicle have been used, together with the generated lateral irregularities. Second, the same simulation by the SM has been carried out, using in this case the equivalent parameters $\mathbf{p}_{opt}$ to be identified, and the generated lateral irregularities. Note that, in order to reduce the complexity of the parameter identification problem, only the lateral alignment,  $\xi$, has been included in simulations with both models, CM and SM, thereby excluding vertical irregularity from the problem. Therefore, by comparison of the dynamic response of the wheelset calculated by both models, the equivalent parameters $\mathbf{p}_{opt}$ can be obtained.

There are several parameter identification methods that can be used to match the dynamic response of the system. In this work, the \emph{Temporal Structural Model Updating Method} \citep{Kraft2012} has been used. This is a time domain approach widely used in different fields, the criterion of which is defined as the difference between the real and modelled time responses. This difference has been evaluated by a misfit function defined in the time domain: the least square error criterion. Using the square of the L2 norm, the cost function can be written as a sum over the channels, at the time step $k$:

\begin{equation} \label{eq:Cost_function}
 J_{ls}(\mathbf{p}_{opt}) =  \frac{1}{N} \sum_{k} \vert \mathbf{x}_{real} (k) -\mathbf{x}_{mod} (k,\mathbf{p}_{opt}) \vert ^{2}
\end{equation}

$\mathbf{x}_{real}$ and $\mathbf{x}_{mod}$ being the state vectors of the real and the modelled system, respectively. In the state vectors, any representative variable can be included. In this work, the most relevant variables in the dynamic behaviour of the vehicle have been chosen: $y$ and $\psi$. Finally, the equivalent parameters of the SM, $\mathbf{p}_{opt}$, are identified by applying a parametric optimisation method which minimises the distance between model and real responses.

\section{Kalman filter} \label{Sec:KF}

The main objective of this work is to estimate the lateral track irregularities from experimental measurement of the dynamic response of the wheelset and the bogie frame: acceleration in the lateral direction ($\ddot{y}$) and yaw angular velocity ($\dot{\psi}$) of the wheelset, and acceleration in the lateral direction ($\ddot{y}^{f}$) of the suspended frame. This estimation is based on the well-known Kalman filter algorithm \cite{Welch2006}.

\subsection{Design of the filter}

The state vector is composed of the generalized coordinates, $\mathbf{q}$, their derivatives, $\mathbf{\dot{q}}$, and the lateral irregularity, $\xi$, as follows:

\begin{equation} \label{eq:State_vector}
\mathbf{x} =\left [ y \; \; \; \psi \; \; \; y^{f} \; \; \;  \dot{y} \; \; \; \dot{\psi} \; \; \; \dot{y}^{f} \; \; \; \xi \right ]^{T}
\end{equation}

The measurement vector is composed of the acceleration and the angular velocity of the wheelset, and the acceleration of the suspended frame, plus an additional measurement of the lateral irregularity: 

\begin{equation} \label{eq:Measurement_vector}
\mathbf{z}_{meas} =\left [ \ddot{y}_{meas} \; \; \; \dot{\psi}_{meas} \; \; \; \ddot{y}^{f}_{meas} \; \; \; \xi_{meas} \right ]^{T}
\end{equation}

It is important to note that the measurement of the lateral irregularity, $\xi_{meas}$, is not a real measurement but a virtual sensor, with zero value, which has been included in the measurement vector with the aim of avoiding a drift in the prediction of the lateral irregularity. 

With the aim of obtaining the equations of the Kalman filter, the equation of motion \eqref{eq:Eq_motion} can be rewritten in the following state-space representation:

\begin{equation} \label{eq:Eq_StateSpace}
\begin{bmatrix}
\mathbf{\dot{q}}\\ 
\mathbf{\ddot{q}}
\end{bmatrix} = 
\begin{bmatrix}
\mathbf{0} & \mathbf{I} \\ 
\mathbf{-M^{-1}\left [ \mathbf{K}_{s} + \mathbf{K}_{c} \right ]}  & 
\mathbf{-M^{-1}\left [ \mathbf{C}_{s} + \mathbf{C}_{c} \right ]} 
\end{bmatrix}
\begin{bmatrix}
\mathbf{q}\\ 
\mathbf{\dot{q}}
\end{bmatrix} + 
\begin{bmatrix}
\mathbf{0}\\ 
\mathbf{M^{-1} Q}_{c,0}(\xi,\dot{\xi})
\end{bmatrix}
\end{equation}

Additionally, a common assumption is to consider the additional state $\xi$ as constant, i.e, its time derivative with a zero value:

\begin{equation} \label{eq:Extra_Estimation_State}
\dot{\xi} = 0
\end{equation}

Assembling Eqs. \eqref{eq:Eq_StateSpace} and \eqref{eq:Extra_Estimation_State}, the system and the measurement equations in the continuous form are given by:

\begin{equation} \label{eq:System_eq}
\mathbf{\dot{x}}(t) = \mathbf{F_{c}} \; \mathbf{x}(t)  + \mathbf{v}(t)
\end{equation}

\begin{equation} \label{eq:Measurement_eq}
\mathbf{z} (t) = \mathbf{H_{c}} \; \mathbf{x}(t) + \mathbf{w}(t)
\end{equation}

where $\mathbf{F_{c}}$ and $\mathbf{H_{c}}$ are the constant state transition and the measurement matrices in the continuous form, respectively, while $\mathbf{v}$ and $\mathbf{w}$ are assumed as Gaussian white noises that can be modelled as: $\mathbf{v}(t) \sim N(0,\mathbf{Q}(t))$ and $\mathbf{w}(t) \sim N(0,\mathbf{R}(t))$, where $\mathbf{Q}(t)$ and $\mathbf{R}(t)$ are the covariance matrices.

The state transition matrix, $\mathbf{F_{c}}$, can be obtained from Eqs. \eqref{eq:Eq_StateSpace} and \eqref{eq:Extra_Estimation_State}:

\begin{equation} \label{eq:Eq_Fc}
\mathbf{F_{c}} = 
\begin{bmatrix}
\mathbf{0} & \mathbf{I} & \mathbf{0} \\ 
\mathbf{-M^{-1}\left [ \mathbf{K}_{s} + \mathbf{K}_{c} \right ]}  & 
\mathbf{-M^{-1}\left [ \mathbf{C}_{s} + \mathbf{C}_{c} \right ]}  &
\mathbf{M^{-1} K_{d}} \\
\mathbf{0} & \mathbf{0} & \mathbf{0} 
\end{bmatrix}
\end{equation}

being:

\begin{equation} \label{eq:Eq_Kd}
\mathbf{K_{d}} = 
\left.\begin{matrix}
\frac{\partial \mathbf{Q}_{c,0} }{\partial \xi} 
\end{matrix}\right| _{\dot{\xi}=0} =
\begin{bmatrix}
2\alpha g (m-m_{f}) / l \\
2\alpha f_{11} l / r_{0} \\
0 \\
\end{bmatrix}
\end{equation}

The measurement matrix, $\mathbf{H_{c}}$, is obtained as:

\begin{equation} \label{eq:Eq_Hc}
\mathbf{H_{c}} = 
\begin{bmatrix}
\multicolumn{7}{c}{[\mathbf{F_{c}}(4,:)]} \\
[ 0 & 0 & 0 & 0 & 1 & 0 & 0 ] \\
\multicolumn{7}{c}{[\mathbf{F_{c}}(6,:)]} \\
[ 0 & 0 & 0 & 0 & 0 & 0 & 1 ]\\
\end{bmatrix}
\end{equation}

In the definition of $\mathbf{H_{c}}$, Matlab-like notation has been used.

The estimator can be implemented in a discrete form, by using a modification of the Euler method in which position integration is discretised using a second order Taylor expansion instead of the standard first order one, leading to the following discrete equations:

\begin{equation} \label{eq:System_discr_eq}
\mathbf{x}_{k} = \mathbf{F} \; \mathbf{x}_{k-1} + \mathbf{v}_{k}
\end{equation}

\begin{equation} \label{eq:Measurement_discr_eq}
\mathbf{z}_{k} =  \mathbf{H} \; \mathbf{x}_{k} + \mathbf{w}_{k}
\end{equation}

where the subscript \textit{k} represents discrete time. In this case,  $\mathbf{F}$ and $\mathbf{H}$ are the constant state transition and the measurement matrices in the discrete form, respectively. 

The Kalman filter is made up of two fundamental steps: estimates and updates. Being ($\hat{\bullet}$) the estimates, the following initial conditions are considered for the state estimates and the error covariance:

\begin{equation} \label{eq:Inicial_x0}
\hat{\mathbf{x}}_{0}^{+} = E[\mathbf{x}_{0}]
\end{equation}

\begin{equation} \label{eq:Inicial_P0}
\mathbf{P}_{0}^{+}=E[ (\mathbf{x}_{0} - \hat{\mathbf{x}}_{0}^{+} )   (\mathbf{x}_{0} - \hat{\mathbf{x}}_{0}^{+} )^{T} ]
\end{equation}

with $E$ the expected value.

The state estimates and the estimation of the error covariance are given by:

\begin{equation} \label{eq:xk_}
\hat{\mathbf{x}}_{k}^{-} = \mathbf{F} \;  \hat{\mathbf{x}}_{k-1}^{+}
\end{equation}

\begin{equation} \label{eq:Pk_}
\mathbf{P}_{k}^{-}= \mathbf{F} \mathbf{P}_{k-1}^{+}  \mathbf{F}^{T} + \mathbf{Q}
\end{equation}

By the computation of the filter gain, $\mathbf{K}_{k}$, and evaluating the measurement residual, the updates of the state estimates and of the estimation of the error covariances can be determined by:

\begin{equation} \label{eq:Kk}
\mathbf{K}_{k}= \mathbf{P}_{k}^{-} \mathbf{H}^{T} ( \mathbf{H} \mathbf{P}_{k}^{-} \mathbf{H}^{T} + \mathbf{R})
\end{equation}

\begin{equation} \label{eq:xk+}
\hat{\mathbf{x}}_{k}^{+}  =  \hat{\mathbf{x}}_{k}^{-}  + \mathbf{K}_{k}\left [ \mathbf{z}_{meas,k}  - \mathbf{H} \; \mathbf{x}_{k}  \right ]
\end{equation}

\begin{equation} \label{eq:Pk+}
\mathbf{P}_{k}^{+}  = (\mathbf{I} -\mathbf{K}_{k}  \mathbf{H}) \mathbf{P}_{k}^{-}
\end{equation}

The performance of the Kalman filter strongly depends on the observability of the system: the system is observable if its behaviour can be determined from output sensors only. For time-invariant linear systems in the state-space representation, there is a convenient test to check whether a system is observable. If the row rank of the following observability matrix:

\begin{equation} \label{eq:Obs_matrix}
\mathbf{O} = \begin{bmatrix}
\mathbf{H} \\ 
\mathbf{H} \mathbf{F} \\ 
\mathbf{H} \mathbf{F}^{2} \\ 
\vdots \\
\mathbf{H}  \mathbf{F}^{n-1}  
\end{bmatrix}
\end{equation}

is equal to $n$ (the number of state variables), then the system is observable. This will be the initial test in the process of estimating the lateral irregularity.

\subsection{Estimation of covariance matrices}

In the Kalman filtering process, a good estimation of the system and measurement covariance matrices ($\mathbf{Q}$ and $\mathbf{R}$) is essential for the good performance of the filter. Both matrices can be estimated from the real system state and measurement vectors, obtained through the CM. Being $\mathbf{x}$ and $\mathbf{z}$ the real system state and measurement vectors, respectively, the covariance matrices can be evaluated. Regarding the system covariance matrix, $\mathbf{Q}$, it depends on how well the system is modelled through the $\mathbf{F}$ matrix. Consequently, for the estimation of $\mathbf{Q}$, it is necessary to first evaluate the system error vector at each time step $k$:

\begin{equation} \label{eq:error_x}
\mathbf{e}^{x}_{k}  = \left [  [\mathbf{x}_{k}]_{CM} -  \mathbf{F} \; [\mathbf{x}_{k-1}]_{CM} \right ]
\end{equation}

where $[\mathbf{x}_{k}]_{CM}$ and $[\mathbf{x}_{k-1}]_{CM}$ are the state vectors evaluated through the CM, at the time step $k$ and $k-1$, respectively. From this, the system covariance matrix is estimated by computing the covariance of the system error, as follows:

\begin{equation} \label{eq:Q}
\mathbf{Q} =  \frac{1}{N} \sum_{k}   \mathbf{e}^{x}_{k} \;\; \mathbf{e}^{x T}_{k} 
\end{equation}

Regarding the measurement covariance matrix, $\mathbf{R}$, it depends on how well the measurement is modelled through the $\mathbf{H}$ matrix and sensors. The measurement error or innovation vector at each time step ($k$) is evaluated as:

\begin{equation} \label{eq:error_z}
\mathbf{e}^{z}_{k}  = \left [  [\mathbf{z}_{k}]_{CM} -  \mathbf{H} \; [\mathbf{x}_{k}]_{CM} \right ]
\end{equation}

where $[\mathbf{x}_{k}]_{CM}$ and $[\mathbf{z}_{k}]_{CM}$ are the state and measurement vectors evaluated through the CM, at the time step $k$. From this, the measurement covariance matrix is estimated by computing the covariance of the measurement error, as follows: 

\begin{equation} \label{eq:R}
\mathbf{R} =  \frac{1}{N} \sum_{k}   \mathbf{e}^{z}_{k} \;\; \mathbf{e}^{z T}_{k} 
\end{equation}

Note that the measurement vector $[\mathbf{z}_{k}]_{CM}$ has been evaluated through the CM and contaminated with the Gaussian white noise of the sensors, with a variance $\boldsymbol{\sigma}_{sensors}$. Consequently, the sensor errors are included in the measurement covariance matrix $\mathbf{R}$.

\section{Results} \label{Sec:Results}

In simulations, the model of the vehicle ML95 operated by the Lisbon subway \citep{Pombo2004} has been used. Geometric and mechanical properties of the vehicle can be found in \cite{Pombo2004}. The case of study is the ML95 vehicle running on a straight track with irregularities (vertical and lateral), at constant forward velocity, $V$ = 20 m/s. A total time of 20 $s$ has been simulated, corresponding to 400 m track length. As previously explained, the Kalman filter needs an estimation of the sensor noise variance,  $\boldsymbol{\sigma}_{sensors}$. The noise variance has been estimated as 10\% of the maximum absolute value of the signals, which is a reasonably realistic working environment for civil engineering applications. Therefore, a value of 0.01 $m/s^2$ for the accelerometers ($\ddot{y}_{meas}$ and $\ddot{y}^{f}_{meas}$) and 0.0005 $rad/s$ for the gyroscope ($\dot{\psi}_{meas}$), has been taken. For the lateral irregularity ($\xi_{meas}$), a value of 5 mm has been fixed, which is the order of the expected value of the lateral irregularity.

\subsection{Generation of track irregularities}

\begin{figure}[h]
	\includegraphics[width=0.95\textwidth,center]{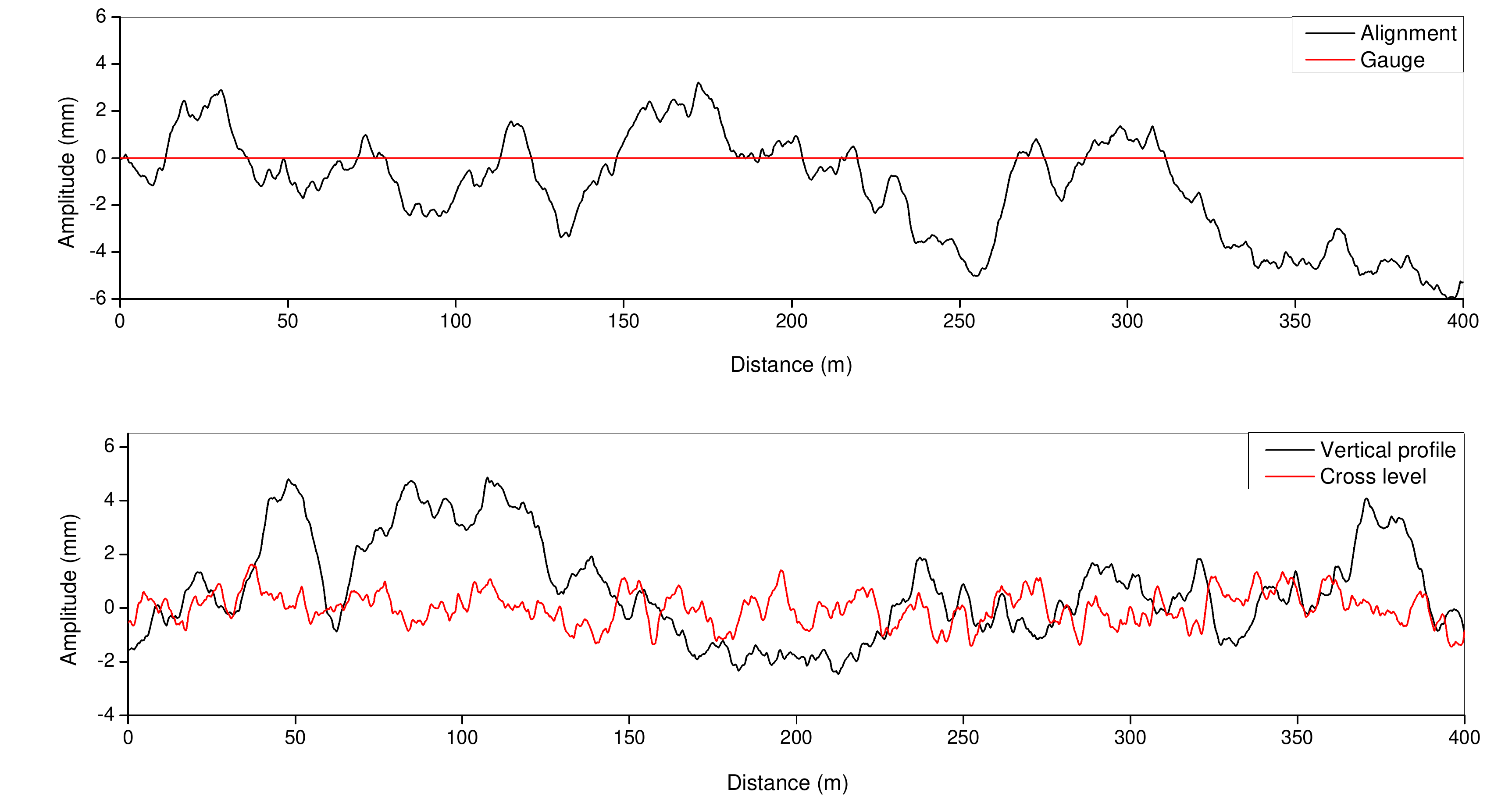}
  	\setlength{\abovecaptionskip}{-20pt}
 	\caption{Lateral and vertical track irregularities}
  	\label{fig:irregularities}
\end{figure}

For the generation of vertical and lateral track irregularities to be included in the models, analytical expressions of the power spectral density functions (PSD) are used. Using the method reported in \citep{CLAUS1998}, vertical and lateral irregularities for both the left and right rails have been generated for a 400 m track length, as shown in Fig. \ref{fig:irregularities}. Recall that gauge is assumed to be constant.

\subsection{Identification of equivalent parameters }

In this section, the identification of the equivalent parameters of the SM has been carried out. Table \ref{tab_parameters} shows the entire set of parameters for the SM, once the parameter identification has been achieved.

\begin{table}[h]
  \begin{center}
    \begin{tabular}{c|l|l|l} 
      \textbf{Parameter} & \textbf{Description} & \textbf{Value} & \textbf{Units}\\
      \hline
      $m$ 		& Wheelset mass 						& 1109 		& Kg \\
      $I$ 		& Wheelset yaw moment of inertia 		& 606  		& Kg.m$^{2}$\\
      $l$ 		& Half width of the wheelset    	 	& 0.75 		& m\\
      $l_{s}$ 	& Length to the primary suspension  	& 0.85 		& m\\
      $\alpha$ 	& Nominal conicity 						& 0.1 		& \\
      $r_{0}$   & Rolling radius of the wheels  		& 0.85 		& m\\
      $m_{f}$   & Mass of the suspended frame		    & 3781      & Kg \\
      $f_{11}$  & Longitudinal creep coeficient         & 5.5e6     & N \\
      $f_{22}$  & Lateral creep coeficient				& 5e6       & N \\
      $f_{23}$  & Spin creep coeficient					& 9.3e3     & N.m\\
      $f_{33}$  & Spin creep coeficient					& 15        & N.m$^{2}$\\
      $k_{x}$   & Longitudinal suspension stiffness     & 7.95e5    & N/m \\
      $c_{x}$   & Longitudinal damper coefficient       & 1.47e4	& N.s/m \\
      $k_{y}$   & Lateral suspension stiffness			& 4.12e6    & N/m \\
      $c_{y}$   & Lateral damper coefficient			& 1.41e5    & N.s/m \\
    \end{tabular}
  \end{center}
      \caption{Equivalent parameters of the SM}
  \label{tab_parameters}
\end{table}

\begin{figure}[h]
	\includegraphics[width=0.95\textwidth,center]{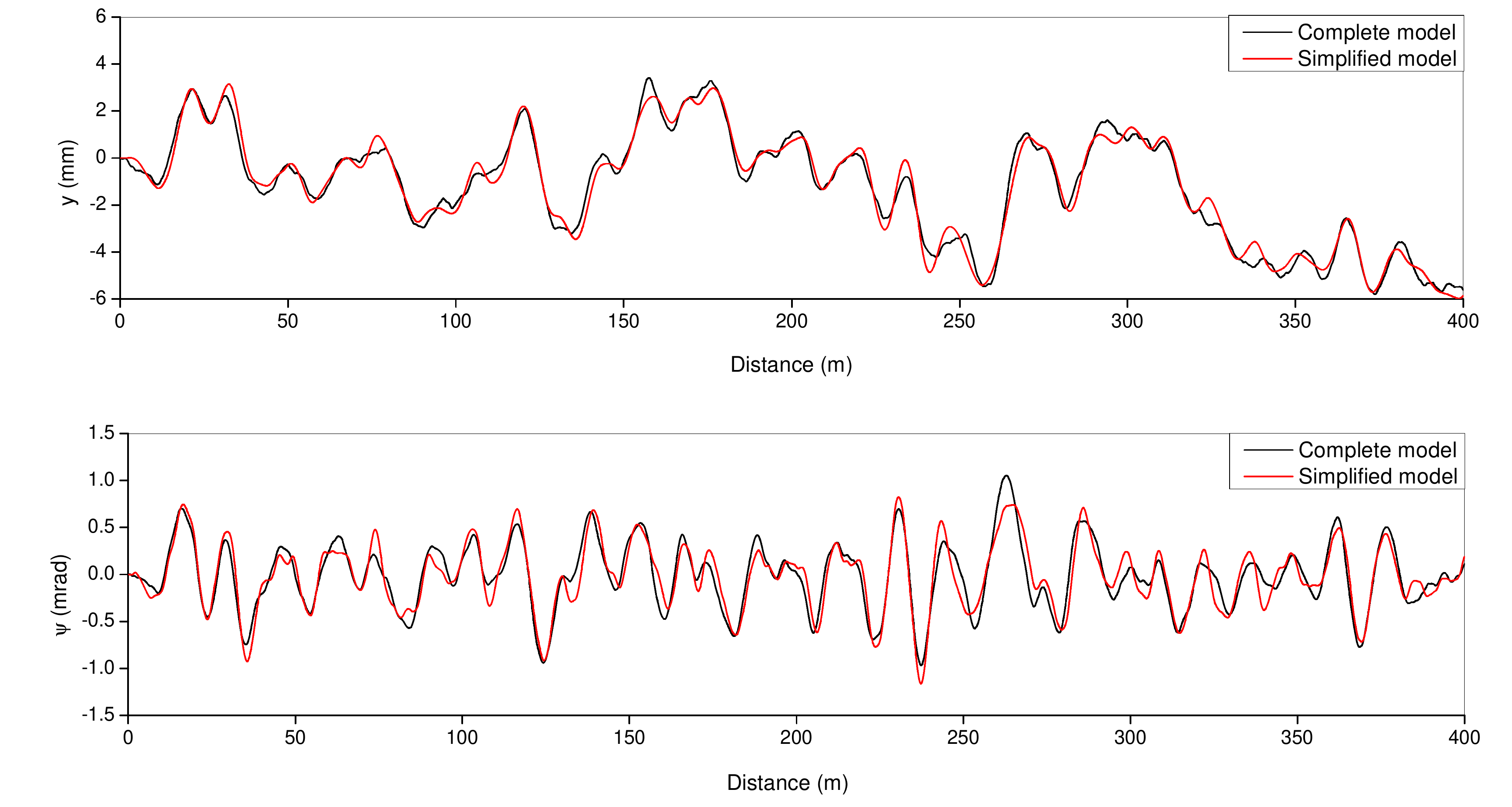}
	\setlength{\abovecaptionskip}{-20pt}
  	\caption{Dynamic models adjustment}
 	 \label{fig:modelsAdjustment}
\end{figure}

Figure \ref{fig:modelsAdjustment} presents the estimation through both models, CM and SM, of the variables included in the state vectors for the optimisation procedure: $y$ and $\psi$. As can be observed in the figure, quite good agreement has been achieved in the simulated dynamic behaviour of both models, taking into account the simplicity of the SM.

\subsection{Estimation of lateral irregularity} \label{SubSec:Irr_estim}

\begin{figure}[h]
	\includegraphics[width=0.95\textwidth,center]{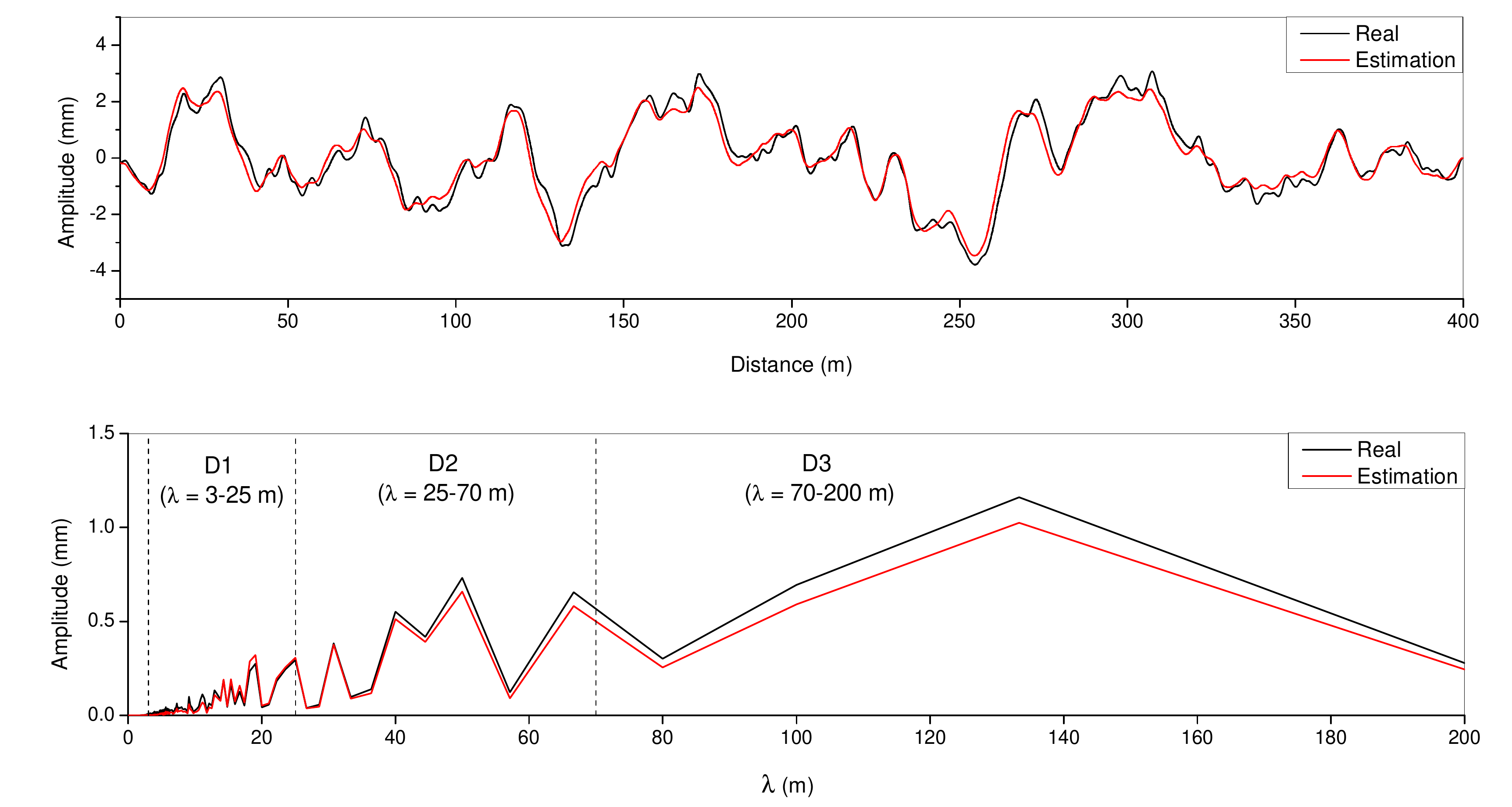}
	\setlength{\abovecaptionskip}{-20pt}
  	\caption{Lateral irregularity estimation filtered in the whole range ($\lambda$ = 3 - 200 m)}
  	\label{fig:Irregularity_estimation}
\end{figure}

Once the equivalent parameters have been estimated, the prediction of the lateral irregularity has been carried out through the proposed Kalman filter algorithm. To this end, the synthetic sensor data have been generated through the CM, taking into account both vertical and lateral irregularities, presented in Fig. \ref{fig:irregularities}. Furthermore, these synthetic measurements have been contaminated with the Gaussian white noise of the sensors, with a variance $\boldsymbol{\sigma}_{sensors}$. The evaluation of the observability matrix, $\mathbf{O}$, in Eq. \ref{eq:Obs_matrix}, with equal rank to the number of state variables ($n$ = 7), confirms that the system is observable. Figure \ref{fig:Irregularity_estimation} shows the comparison between the estimated lateral track irregularity and the reference one.

In the upper subplot, the two irregularity profiles are compared in the space domain, for the 400 m track-length under study, whereas the lower subplot shows both profiles in the frequency domain, obtained by the FFT. It should be noted that both profiles, estimated and real, have been filtered with a Butterworth bandpass filter in the range of interest, according to the standards \cite{Standars_EN13848}: frequencies corresponding to a wavelength between 3 and 200 m. In light of the results, a good agreement of the estimated and the real lateral irregularity is observed across the entire length of the spatial profile. The results in the frequency domain complement the information obtained, showing a good prediction of the lateral irregularity throughout the whole frequency range, which is divided into three ranges according to the standards: D1 ($\lambda$ = 3-25 m),  D2 ($\lambda$ = 25-70 m) and D3 ($\lambda$ = 70-200 m).

\begin{figure}[h]
	\includegraphics[width=0.95\textwidth,center]{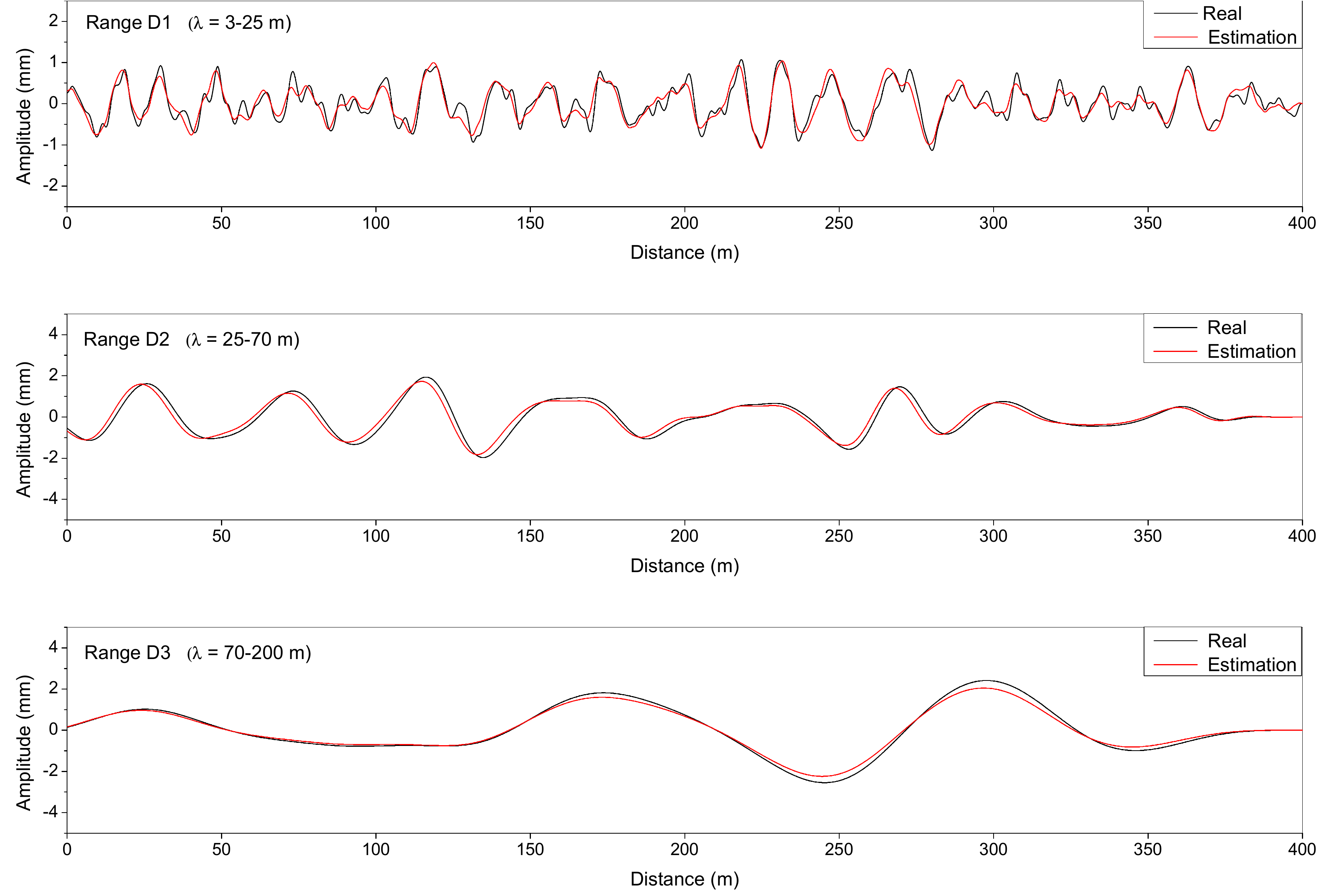}
	\setlength{\abovecaptionskip}{-20pt}
  	\caption{Lateral irregularity estimation filtered in different ranges: D1 ($\lambda$ = 3-25 m), D2 ($\lambda$ = 25-70 m) and D3 ($\lambda$ = 70-200 m)}
  	\label{fig:Irregularity_estimation_D123}
\end{figure}

For a more in-depth analysis, the results obtained have to be divided into the three different ranges, by filtering them into the corresponding bandpass limits (i.e. D1, D2 and D3). Therefore, results have been plotted in Fig. \ref{fig:Irregularity_estimation_D123}, where the comparison between estimated and real lateral irregularity are presented in the three different ranges, in the space domain. Again, very good agreement is obtained in the three wavelength ranges.

In order to numerically evaluate the results achieved with the proposed Kalman filter estimator, an accuracy index has been calculated. In this work, two different accuracy indices have been used:

\begin{equation} \label{eq:Accuracy_index}
\begin{split}
&  J = rms \; (\mathbf{\xi}_{est} - \mathbf{\xi}_{real}) \\
\\
&  J_{rel} =  \frac {rms \; (\mathbf{\xi}_{est} - \mathbf{\xi}_{real})}  {rms \; (\mathbf{\xi}_{real})}    
\end{split}
\end{equation}

The first one is the absolute accuracy index $J$, calculated as the root mean square value (rms) of the difference between the estimated and the real lateral irregularity. This index $J$ has length units and is particularly useful and intuitive for measuring the disagreement of the estimation with the real data. The second one is the relative accuracy index $J_{rel}$, which corresponds to the non-dimensional value of $J$. This index has no dimensions and completes the information of the absolute accuracy index. Therefore, the accuracy indices for the estimations, according to different wavelength ranges, are shown in Table \ref{tab_Results_KF}. It can be seen that values of $J$ = 0.36 mm and $J_{rel}$ = 0.25 are achieved in the estimation when the whole spectrum is considered, confirming the accuracy of the estimator. It is important to note that, in the standard case analysed in this section, with the aim of being realistic, the vertical irregularities of the track and the sensor noise have been included in the synthetic measurement sensor data. Both factors are a source of errors in the prediction of the lateral irregularities, making the Kalman filter estimator process more difficult. Nevertheless, very good results are obtained in the estimations.
When analysing the different wavelength ranges, a considerable reduction of the absolute accuracy index $J$ is obtained, particularly in the D3 range. It must be noted that the relative accuracy index $J_{rel}$ is significantly higher in the D1 range. This fact is explained by the lower magnitude value of the D1 irregularities (see Fig. \ref{fig:Irregularity_estimation_D123}).

\begin{table}[h]
  \begin{center}
    \begin{tabular}{c|c|c|c} 
      Whole range & D1 range & D2 range & D3 range\\
      \hline
      0.36 / 0.25   &   0.21 / 0.48  &   0.22 / 0.26  & 0.18 / 0.15 \\
	\hline
    \end{tabular}
  \end{center}
      \caption{Accuracy indices, $J$ (in mm) / $J_{rel}$, in different wavelength ranges}
  \label{tab_Results_KF}     
\end{table}

Finally, in order to test the efficiency of the proposed Kalman filter, the computing time to simulate the case under study has been calculated. The algorithm has been developed in Matlab R2016a with a computer with an Intel Core i7 CPU 2600 3.4 GHz processor. Only 5.7 $s$ of computation time has been required to simulate the total time of the case under study, 20 $s$. This number can even be improved significantly if the Kalman filter is implemented using a low-level programming language like Fortran or C/C++. Consequently, the proposed algorithm is particularly appropriate for real-time applications.

\subsection{Robustness to parameter uncertainty}

A numerical analysis has been performed, varying the uncertain parameters that could change with the running conditions to evaluate their effect on the estimation. The summary of the obtained results can be seen in Table \ref{tab_Sensitivity}, where different conditions are evaluated and compared with the standard one, which was previously analysed. For each case, only one effect is analysed at a time.

\begin{table}[h]
  \begin{center}
    \begin{tabular}{l|c|c|c|c} 
     \textbf{Condition}  & Whole range & D1 range & D2 range & D3 range\\
      \hline
			Standard 					& 0.36 / 0.25   &   0.21 / 0.48  &   0.22 / 0.26  & 0.18 / 0.15 \\
	  \hline  
	  		 Without sensor noise 		& 0.34 / 0.24  &   0.21 / 0.48  &   0.21 / 0.26  & 0.13 / 0.12 \\
	  \hline  
	 		 With no vertical irreg. 	& 0.34 / 0.24 &   0.17 / 0.40  &   0.21 / 0.26 & 0.18 / 0.16 \\
      \hline  
	 		 Conicity ($-10\%$) 		& 0.38 / 0.27 &   0.22 / 0.51 &   0.23 / 0.29 & 0.18 / 0.16 \\		 	       								     
      \hline  
	  		 Kalker's coefficients ($-50\%$) & 0.39 / 0.27 &   0.22 / 0.51 &   0.23 / 0.29 & 0.18 / 0.16 \\	
      \hline  
	  		 All conditions together & 0.47 / 0.32 &   0.30 / 0.68 &   0.27 / 0.32 & 0.21 / 0.18\\	     					 	       					    	
	\hline
    \end{tabular}
  \end{center}
      \caption{Accuracy indices, $J$ (in mm) / $J_{rel}$, in different wavelength ranges, for different conditions}
  \label{tab_Sensitivity}
\end{table}

First, the effect of the measurement noise on the predictions has been evaluated. To this end, the synthetic measurement generated through the CM to be used in the Kalman filter has been used without noise. With these noise-free measurements, the estimation has been performed and the accuracy indices evaluated. It can be seen that, as expected, there is an increase of the estimation accuracy. However, the effect of the sensor noise is not relevant.

Second, the effect of the vertical irregularities on the estimation of the lateral irregularity has been analysed. In this case, the generation of the synthetic measurement has been performed through the CM, including only the lateral irregularities shown in Fig. \ref{fig:irregularities}, without vertical ones. With these new synthetic measurements, the estimation has been performed and the accuracy indices evaluated and included in Table \ref{tab_Sensitivity}. Obviously, an improvement in the estimation is achieved. However, it can be concluded that the vertical irregularities hardly affect the estimations, as could be expected: the vertical irregularities being much smaller than the width of the wheelset, it hardly affects the dynamics of the wheelset and, consequently, the estimation results.

Third, the effect of the uncertainty in the wheel conicity on the estimations has been studied. This parameter is especially important for different reasons. First, the wheels of a railway vehicle are not usually conical, as it has been assumed in this work. Second, the profile of the wheels can change throughout their life, due to the wear from contact with the rails. Consequently, the value of the conicity cannot be accurately found out and therefore has some degree of uncertainty. In this analysis, a variation in the conicity value of $-10\%$ has been included in the Kalman filter and in the results in the simulations evaluated through the accuracy indices. A mild effect on the estimation is observed in the results.

Fourth, the contact conditions between the wheels and the rails have been considered. In order to evaluate the effect of the uncertainty in the Kalker’s coefficients, the synthetic measurements have been generated using the CM, but by reducing the Kalker’s coefficients to 50\%. With these new synthetic measurements, the estimation has been carried out and the accuracy indices of the results evaluated. As observed, the accuracy in the prediction of the lateral irregularity under this new condition decreases, although slightly.

Finally, the worst-case scenarios have been considered, i.e., all adverse conditions together at the same time (noise, vertical irregularities, conicity uncertainty and Kalker’s coefficients uncertainty). In this critical case, the results in the prediction are quite acceptable, with just a moderate increase of the accuracy indices compared with the standard case.

Consequently, after considering the results obtained in the numerical analysis, summarised in Table \ref{tab_Sensitivity}, it can be assumed that among all the parameters that could introduce any kind of uncertainty into the Kalman filter, none has a significant impact in the predictions, even in a critical case in which all conditions are considered.

\subsection{Robustness under resonance conditions}

In order to verify the robustness of the estimator, it has to be checked under very critical, although not likely, conditions. Since the railway vehicle is a mechanical system with its own modes of vibrations, if one of those modes were excited during the ride by track irregularities, the natural movement of the wheelsets would be amplified due to resonance. This amplification leads to higher levels of acceleration and angular velocities in the wheelset and the suspended frame, which are inputs in the Kalman filter estimator, thereby complicating the efficient performance of the estimator. Resonance should not be interpreted as higher levels of lateral irregularities. Due to the fact that the estimator is based on a dynamic model (the SM), it should be able to estimate the real value of irregularities from the lateral motion of the wheelset, even if this movement is amplified by excitation of any mode of vibration.

\begin{figure}[h]
	\includegraphics[width=0.95\textwidth,center]{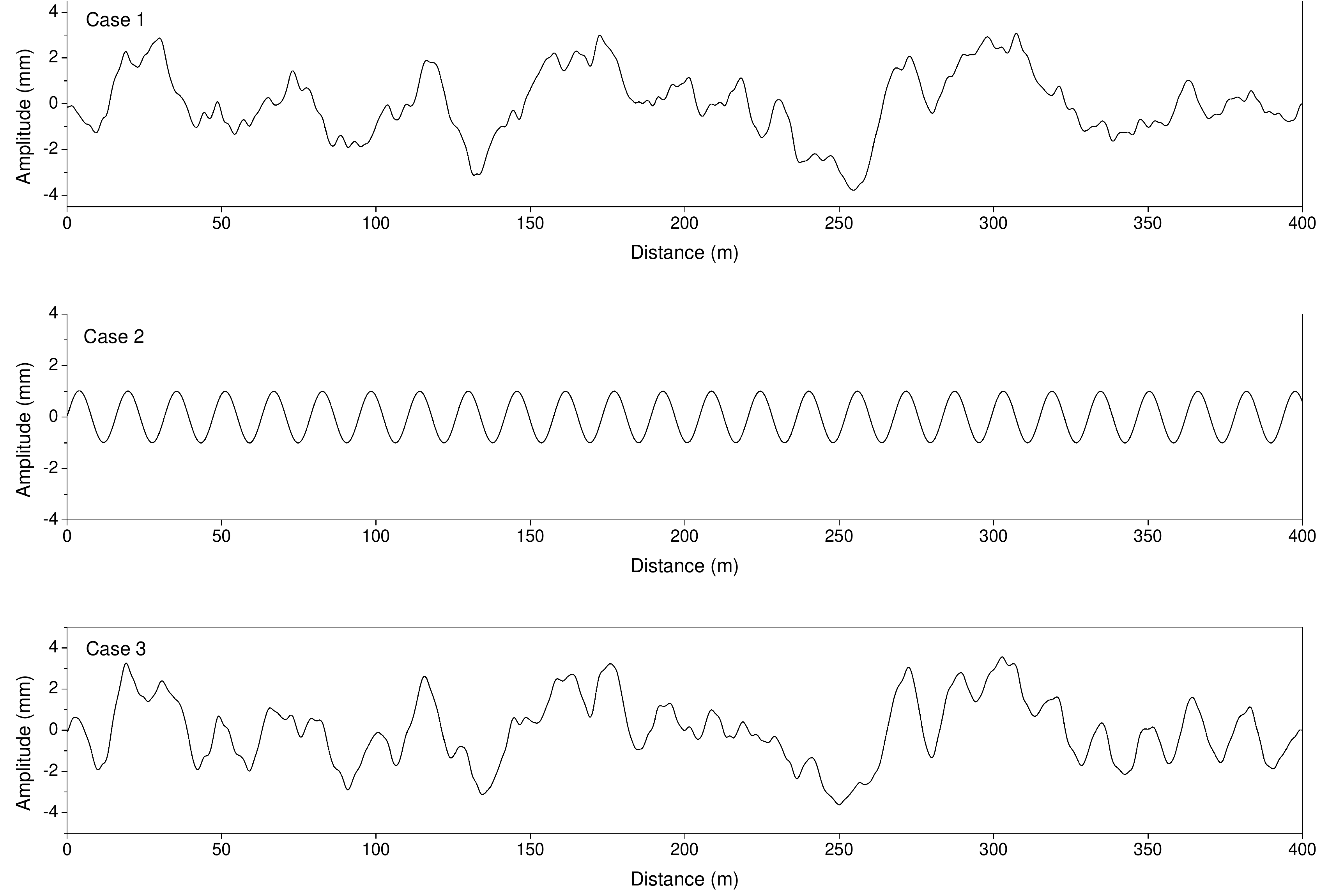}
	\setlength{\abovecaptionskip}{-20pt}
  	\caption{Lateral irregularity for the different cases under study}
  	\label{fig:Excited_Irreg}
\end{figure}

First, the modes of vibration of the vehicle are calculated using modal analysis, performed through the eigenanalysis of the system. The first mode of vibration corresponds to a frequency of 1.277 Hz, which, at a constant velocity of 20 m/s, leads to a wavelength of 15.66 m. In order to validate the Kalman filter, three different cases have been analysed, all of them corresponding to the same vehicle and conditions previously studied. In all cases, the synthetic sensor data have been generated through the CM and contaminated with sensor noise, including vertical irregularities presented in Fig. \ref{fig:irregularities}, but using different lateral irregularities as inputs (see Fig. \ref{fig:Excited_Irreg}). The first case (Case 1) is the standard case previously analysed in Section \ref{SubSec:Irr_estim}. The second one (Case 2) corresponds to the critical case in which the lateral irregularity is a harmonic signal of 1 mm of amplitude and a wavelength $\lambda$ = 15.66 m which, for a constant velocity of 20 m/s, corresponds to a frequency of 1.277 Hz (first natural frequency of the vehicle). This is the critical case in which the harmonic irregularity excites the first mode of vibration. Finally, in the third case of study (Case 3), the lateral irregularity is the sum of the irregularities of cases 1 and 2.

\begin{figure}[h]
	\includegraphics[width=0.95\textwidth,center]{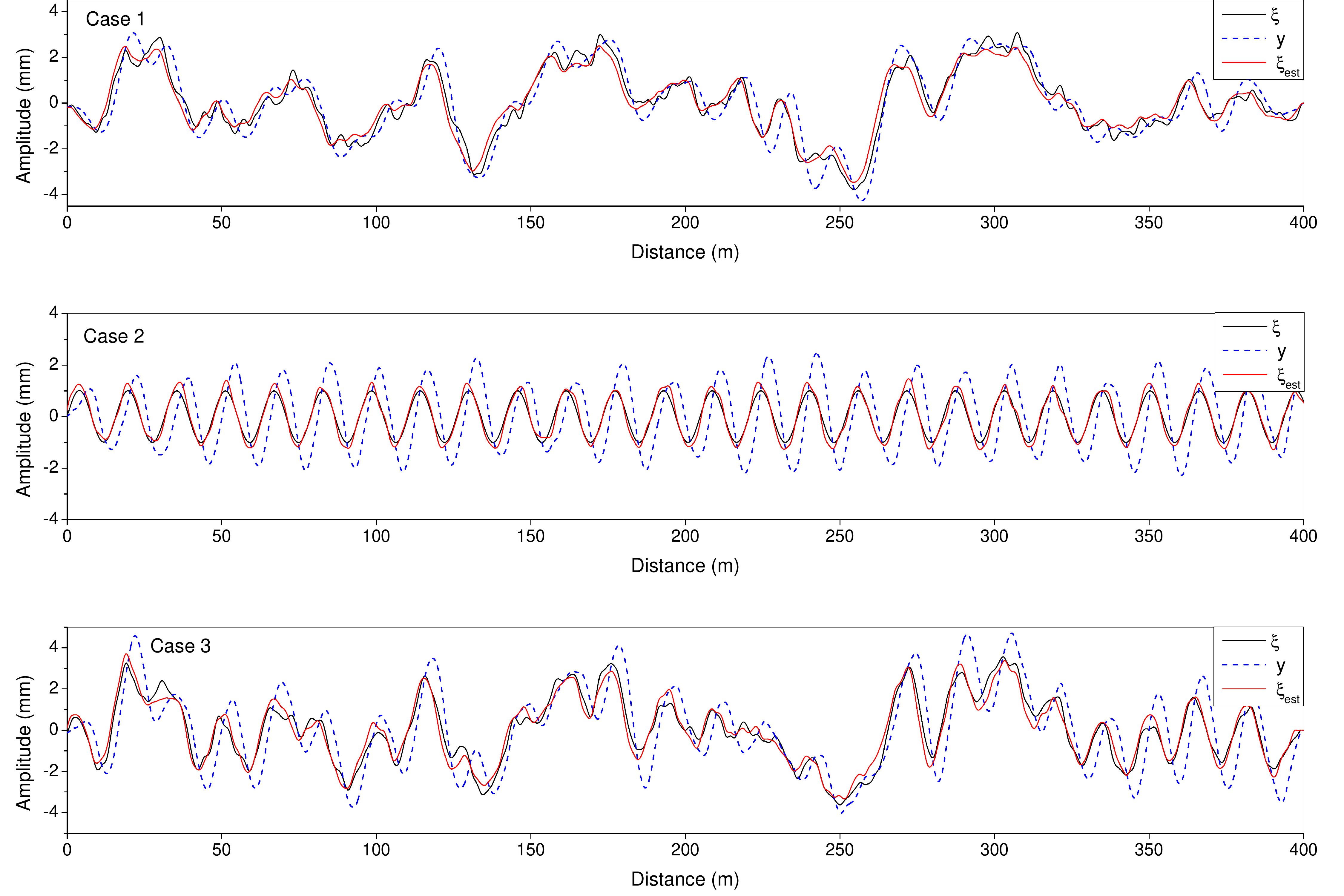}
	\setlength{\abovecaptionskip}{-20pt}
  	\caption{Lateral irregularity estimation for the different cases under study}
  	\label{fig:Irr_Excited_estimation}
\end{figure}

By using the irregularity corresponding to each case, the Kalman filter has been used following the same procedure explained in previous sections. The results of estimations are presented in Fig. \ref{fig:Irr_Excited_estimation}, for the three cases under study. Additionally, with the aim of analysing the performance of the Kalman filter, the lateral displacement of the wheelset has also been plotted. To complete the information, the accuracy indices, $J$ and $J_{rel}$, have been calculated in each case and presented in Table \ref{tab_Reliability}.

\begin{table}[h]
  \begin{center}
    \begin{tabular}{l|c|c|c|c} 
     \textbf{Condition}  & Whole range & D1 range & D2 range & D3 range\\
      \hline
      Case 1 					 	  & 0.36 / 0.25   &   0.21 / 0.48  &   0.22 / 0.26  & 0.18 / 0.15 \\
	 \hline  
      Case 2 					 	  & 0.21 / 0.29  &   0.17 / 0.27  &   0.04 / 0.99  & 0.08 / 5.9 \\
	  \hline      
      Case 3 					 	  & 0.38 / 0.24  &   0.24 / 0.32  &   0.21 / 0.27  & 0.17 / 0.15 \\
	  \hline       
    \end{tabular}
  \end{center}
      \caption{Accuracy indices, $J$ (in mm) / $J_{rel}$, in different wavelength ranges, for different cases under study}
  \label{tab_Reliability}
\end{table}

In the first case, the standard case has been studied. From the results presented in the first plot of Fig. \ref{fig:Irr_Excited_estimation}, it can be appreciated that the wheelset follows the lateral irregularities quite faithfully: the lateral displacement of the wheelset, $y$, is quite similar to the real irregularity, $\xi_{real}$, but with a certain phase delay and some kind of over-oscillations around the peaks of the signal. However, the Kalman filter estimation, $\xi_{est}$, corrects both the phase delay and the over-oscillations, verifying the good performance of the Kalman filter.

In the second scenario (Case 2), the vehicle is excited by the harmonic irregularity at the first natural frequency, amplifying in this case the lateral displacement of the wheelset. The second plot of Fig. \ref{fig:Irr_Excited_estimation} shows the lateral displacement of the wheelset, $y$, to be out of phase and the resulting significant amplification with regard to the input irregularity. In this case, the amplitude ratio (relationship $y/\xi$) is around 2. Again, the Kalman filter makes a very good prediction of the lateral irregularity in this critical case, as can be observed in the figure. These results are corroborated by the accuracy indices obtained in this case (see Table \ref{tab_Reliability}): $J$ = 0.21 mm and $J_{rel}$ = 0.29. Note that, in Case 2, almost the entire absolute error is contained in D1 range, due to the fact that the irregularity is a harmonic signal with a wavelength A = 15.66 m, belonging to the D1 range ($\lambda$ = 3-25 m). In ranges D2 and D3, the absolute error ($J$) is very low, although obviously the relative error ($J_{rel}$) is significant.

Finally, in the last case study (Case 3), a combination of the irregularities of the two previous cases has been taken as input in the Kalman filter. The third plot of Fig. \ref{fig:Irr_Excited_estimation} shows the results in Case 3: similarly to the previous case, the lateral displacement of the wheelset,  $y$, is out of phase and has significant amplification with regard to the input irregularity. This result could be expected, since lateral irregularity in this case has a frequency content corresponding to the first natural frequency of the vehicle, amplifying the lateral motion of the vehicle. Regarding the estimation of the lateral irregularity in this third case, very good results are shown in Fig. \ref{fig:Irr_Excited_estimation} and in Table \ref{tab_Reliability}.

In conclusion, the results obtained in the different cases studied in this section prove that the Kalman filter estimator is quite efficient and robust even in the critical case in which irregularities produce vehicle resonance.

\section{Conclusions and future works} \label{Sec:Conclusions}

In this work, a simple and robust measuring system combined with a dynamic model-based Kalman filter estimator has been proposed to be used on in-service vehicles for continuous monitoring of track geometry and estimation of lateral alignment. The proposed numeric technique is based on the Kalman filtering method, using the measurement from only three inertial sensors (two accelerometers and one gyroscope) mounted on an in-service vehicle, running on a straight track with irregularities.

The Kalman filtering method used is based on a \textit{Simplified Model} (SM) that adequately reproduces the lateral dynamic behaviour of the vehicle, the simplicity of which drastically reduces the computational load of the estimator. The main contribution of the presented work is the use of such a simplified linear dynamic model to be used to perform a classical linear Kalman filter. Consequently, in order to obtain good performance in the proposed estimator, accurate identification of the equivalent parameters for the SM is essential. Otherwise, without a well-characterised SM, the proposed estimator will not be able to provide a good estimation of the lateral alignment. To this end, a parametric optimization method has been used, with very good results, taking into account the simplicity of the SM. To validate the proposed method, virtual experimental data to be used as an input in the Kalman filter have been generated through the \textit{Complete Model} (CM), a detailed dynamic model previously proposed by the authors.

Through use of the proposed method, the result obtained has been analysed in the different wavelength ranges defined in the standards, showing very good agreement in all of them, with maximum errors around 0.3 – 0.4 $mm$. Additionally, the efficiency of the proposed estimator has been checked, showing a very low computational cost, which makes it especially appropriate for real-time applications. Finally, the work has been completed by performing a systematic parametric analysis of the Kalman filter, analysing the influence that the uncertainty of different parameters and running conditions (sensor noise, vertical irregularity, conicity uncertainty and Kalker's coefficients uncertainty) can have on the results of the estimation. In light of the results obtained, the estimator has shown great robustness and reliability. Furthermore, the robustness of the method has been tested in a very critical case in which irregularities produce vehicle resonance.

Based on the results presented, it can be concluded that the proposed methodology (measuring system and model-based Kalman filter estimator) achieves a good compromise between simplicity and precision. Consequently, it is suitable for use on in-service vehicles for continuous monitoring of track condition and for the identification of the lateral alignment of the tracks, and it can also be used in real-time applications. In future work, the assessment of the proposed technique should be experimentally validated, using measurements performed on real in-service vehicles and verifying the accuracy and reliability of the estimator.

\section*{Acknowledgements}

This research was supported by the Spanish Ministry of Economy and Competitiveness under the project reference TRA2017-86355-C2-1-R. This support is gratefully acknowledged.

\section*{References}

\bibliography{Bibio_Tren}



\end{document}